\documentclass[journal=nalefd,manuscript=letter]{achemso}

\usepackage[version=3]{mhchem} 
\usepackage[T1]{fontenc}       

\author{Daniel Weber}
\affiliation[Max Planck Institute for Solid State Research, Heisenbergstr. 1, 70569 Stuttgart, Germany]
{Max Planck Institute for Solid State Research Stuttgart}
\alsoaffiliation[Ludwig Maximilians University, Butenandtstr. 5-13, 81377 Munich, Germany]
{Ludwig Maximilians University Munich}
\author{Leslie M. Schoop}
\affiliation[Max Planck Institute for Solid State Research, Heisenbergstr. 1, 70569 Stuttgart, Germany]
{Max Planck Institute for Solid State Research Stuttgart}
\author{Viola Duppel}
\affiliation[Max Planck Institute for Solid State Research, Heisenbergstr. 1, 70569 Stuttgart, Germany]
{Max Planck Institute for Solid State Research Stuttgart}
\author{Judith M. Lippmann}
\affiliation[Max Planck Institute for Solid State Research, Heisenbergstr. 1, 70569 Stuttgart, Germany]
{Max Planck Institute for Solid State Research Stuttgart}
\alsoaffiliation[Ludwig Maximilians University, Butenandtstr. 5-13, 81377 Munich, Germany]
{Ludwig Maximilians University Munich}
\author{J\"urgen Nuss}
\affiliation[Max Planck Institute for Solid State Research, Heisenbergstr. 1, 70569 Stuttgart, Germany]
{Max Planck Institute for Solid State Research Stuttgart}
\author{Bettina V. Lotsch}
\affiliation[Max Planck Institute for Solid State Research, Heisenbergstr. 1, 70569 Stuttgart, Germany]
{Max Planck Institute for Solid State Research Stuttgart}
\alsoaffiliation[Ludwig Maximilians University, Butenandtstr. 5-13, 81377 Munich, Germany]
{Ludwig Maximilians University Munich}
\alsoaffiliation[Nanosystems Initiative Munich, Schellingstr. 4, 80799 Munich, Germany]
{Nanosystems Initiative Munich}
\email{b.lotsch@fkf.mpg.de}

\title[Magnetic properties of restacked 2D spin~$\frac{1}{2}$ honeycomb RuCl$_3$ nanosheets]
  {Magnetic properties of restacked 2D spin~$\frac{1}{2}$ honeycomb RuCl$_3$ nanosheets}

\begin{document}

\begin{abstract}

Spin~$\frac{1}{2}$ honeycomb materials have gained substantial interest due to their exotic magnetism and possible application in quantum computing. However, in all current materials out-of-plane interactions are interfering with the in-plane order, hence a true 2D magnetic honeycomb system is still of demand. Here, we report the exfoliation of the magnetic semiconductor $\alpha$-\ce{RuCl3} into the first halide monolayers and the magnetic characterization of the spin~$\frac{1}{2}$ honeycomb arrangement of turbostratically stacked \ce{RuCl3} monolayers. The exfoliation is based on a reductive lithiation/hydration approach, which gives rise to a loss of cooperative magnetism due to the disruption of the spin~$\frac{1}{2}$ state by electron injection into the layers. After an oxidative treatment, cooperative magnetism similar to the bulk is restored. The oxidized pellets of restacked single layers feature a magnetic transition at T$_N$~=~7~K in the in-plane direction, while the magnetic properties in the out-of-plane direction vastly differ from bulk $\alpha$-\ce{RuCl3}. The macroscopic pellets of \ce{RuCl3} therefore behave like a stack of monolayers without any symmetry relation in the stacking direction. The deliberate introduction of turbostratic disorder to manipulate the spin structure of \ce{RuCl3} is of interest for research in frustrated magnetism and complex magnetic order as predicted by the Kitaev-Heisenberg model.

\end{abstract}

\vspace{\baselineskip}

The emergence of graphene\cite{graphene,graphene-dirac-e-} initiated the development of a variety of single layer compounds as well as investigations into their electronic, optical and mechanical properties. The materials that are most actively examined are monolayers either composed of a single element, such as carbon based graphene or black phosphorus single layers, or binary transition metal chalcogenides \cite{swell-exfol,2D-review-miro,2D-photonics,2D-electronics}. Binary halide nanosheets have been predicted based on chemical intuition \cite{swell-exfol,2D-review-miro} or \textit{ab initio} calculations \cite{CrCl3-sheet-theory}. Yet, no single layer halides have been synthesized so far, even though this class of compounds features an array of interesting electrical and magnetic properties.

The magnetic semiconductor $\alpha$-\ce{RuCl3} is one such example. While it was investigated in the past as a host for intercalants \cite{RuCl3-intercal-1983,RuCl3-intercal-1986} and as a lithium ion conductor \cite{RuCl3-intercal-Li+polymer}, current research focuses on its magnetic properties. Due to its layered honeycomb structure of spin~$\frac{1}{2}$ \ce{Ru^{3+}} centers in combination with spin orbit coupling (SOC), it is one of the few known materials featuring a zigzag antiferromagnetic (AF) ground state below a temperature of T$_{N1}$~=~8~K\cite{RuCl3-zigzag-exp,RuCl3-mag+NMR,RuCl3-XY-paper}. In the zigzag order, the magnetic moments form ferromagnetic (FM) zigzag chains, whose magnetization direction is opposed to the neighboring chains within the plane. Additionally, there is a further magnetic phase transition observed at T$_{N2}$~=~14~K. The origin of this transition is currently still under debate.

This type of ordering was first observed in \ce{Na2IrO3}\cite{na2iro3-zigzag-exp1,na2iro3-zigzag-exp2,na2iro3-zigzag-exp3} and explained by the Kitaev-Heisenberg model\cite{kitaev,na2iro3-zigzag}, which describes that a frustrated spin~$\frac{1}{2}$ honeycomb arrangement could lead to a variety of interesting spin structures. Based on the competition among the exchange interactions up to the third neighbor, the system could possibly be pushed into a quantum spin liquid regime by the manipulation of the competing interactions, thereby opening up applications in quantum computing\cite{kitaev,quantumcomputing}. Yet, the \ce{Na+} ions in the interlayer space of \ce{Na2IrO3} lead to disadvantageous interactions between the iridate layers, which interfere with theoretical predictions of a honeycomb arrangement of spin~$\frac{1}{2}$ magnetic arrays\cite{RuCl3-theory-kitaev}. Eliminating the interlayer interaction could provide a route to manipulate the spin structure of real materials featuring a spin~$\frac{1}{2}$ honeycomb arrangement.

In \ce{RuCl3}, where no charged ions are in between the honeycomb layers, the interlayer interactions are smaller than in \ce{Na2IrO3}, but are not neglegible\cite{RuCl3-interlayer}. The appearance of several magnetic transitions between the zigzag order at T$_{N1}$~=~8~K and the second ordering temperature of T$_{N2}$~=~14~K have been discussed in regards to the stacking behavior of $\alpha$-\ce{RuCl3}. One argument attributes the transitions to the interplay between the first, second and third nearest neighbor as well as interlayer interactions\cite{RuCl3-XY-paper}, while others claim that ABAB~stacking faults in the ABC stack of \ce{RuCl3} layers are responsible for the variety of magnetic transitions\cite{RuCl3-QSL-proximate}. While both of these arguments emphasize the importance of an ordered stacking, the influence of the interlayer interactions could also be investigated by deliberately decoupling the interlayer interactions. 

Two strategies to further reduce the interlayer interactions and thereby possibly decouple the magnetic behavior between the layers can be conceived. The first is to increase the interlayer space by introducting a nonmagnetic compound. This appproach has been realizied in \ce{Na3Ni2SbO6} and \ce{Na3Ni2BiO_{6-$\delta$}}, where hydration of the interlayer \ce{Na+} ions increases the interlayer distance by about one {\AA}ngstrom\cite{nani2sbo6-hydrate,nani2bio6-hydrate}. Another example is the \ce{CrCl3}-graphite intercalation compound \ce{C_{20.9}CrCl3}, where the intercalation of \ce{CrCl3} single layers into graphite resulted in the emergence of spin glass behavior \cite{CrCl3-graphite}. Exfoliation of 2D~honeycomb compounds into monolayers and restacking the layers to form a solid with turbostratic disorder is the second strategy to reduce interlayer coupling. This kind of disorder can be described by random rotations and/or translations of the sheets around an axis perpendicular to the monolayer. Recently, it was shown that the stacking angle of two \ce{MoS2} monolayers has a strong effect on the electronic interlayer coupling.\cite{Huang2014} Thus, the restacking of single layers into a turbostratically disordered solid should significantly alter the magnetic properties of a layered compounds.

Here, we present a form of \ce{RuCl3} with turbostratic disorder (t-\ce{RuCl3}) introduced by exfoliation into single layers and subsequent restacking into a lamellar, disordered solid. To the best of our knowledge, this is the first report of a single layer halide as well as of a single layer compound with a S~=~$\frac{1}{2}$ honeycomb order. Furthermore, we show the disappearance of cooperative magnetism in the intermediary compound \ce{Li_{0.2}RuCl3} and in the restacked pellet of the partially reduced $[$\ce{RuCl3}$]$\textsuperscript{x-} single layers. By an oxidative reaction step, a multisheet stack of $[$\ce{RuCl3}$]$\textsuperscript{x-} is converted to t-\ce{RuCl3}, which features a magnetic transition at T$_N$~=~7~K. The similarities and differences of the magnetic properties between bulk $\alpha$-\ce{RuCl3} and restacked t-\ce{RuCl3} will be discussed in regards to the changes of the electronic and crystal structure.

\begin{figure}[thbp]
	\centering
		\includegraphics[width=0.9\textwidth]{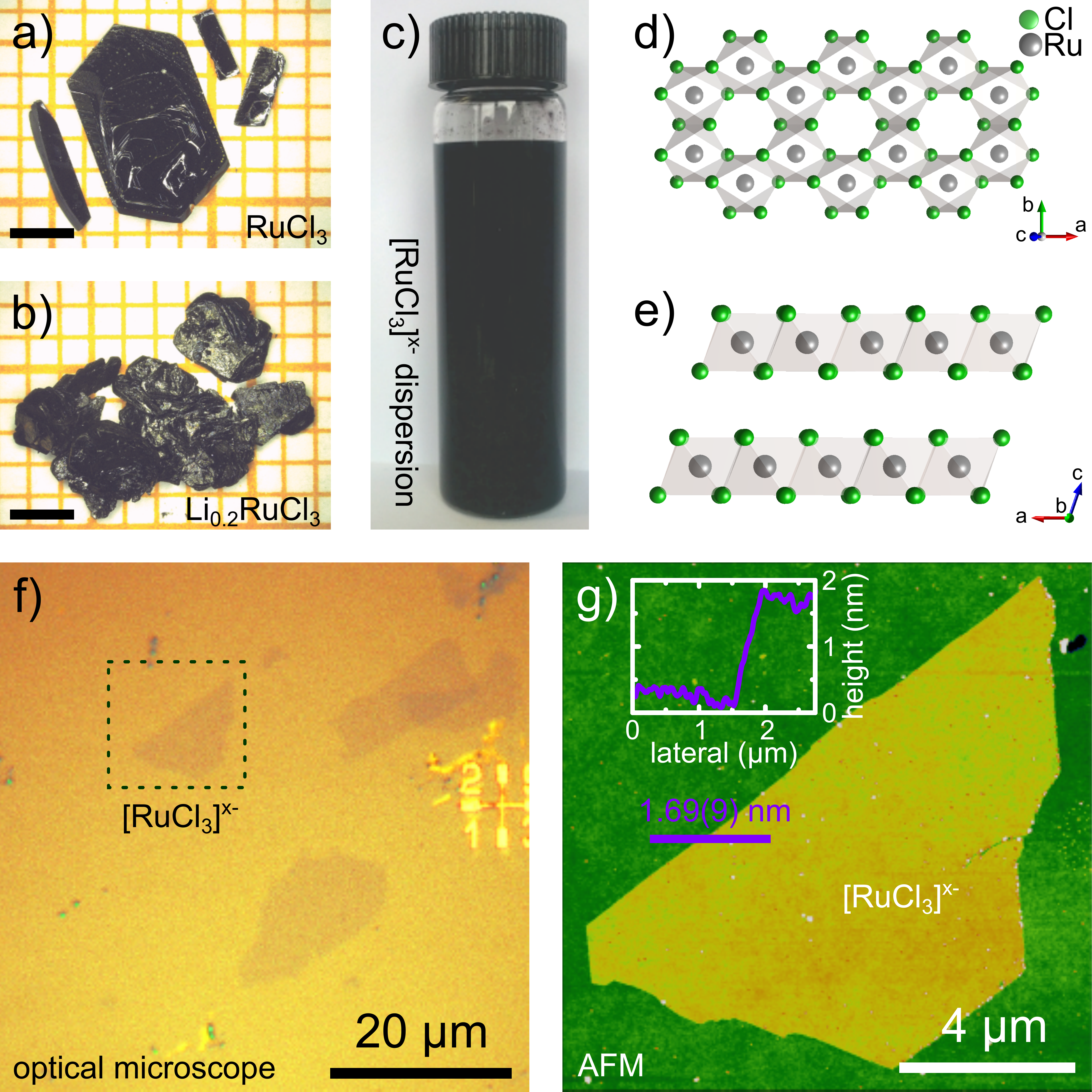}
		\caption{a) $\alpha$-\ce{RuCl3} crystals, scale bar 2 mm, b) \ce{Li_{0.2}RuCl3} platelets, scale bar 2 mm, c) [\ce{RuCl3}]\textsuperscript{x-} nanosheet in aqueous dispersion, d) honeycomb structure of $\alpha$-\ce{RuCl3} viewed along [001], e) layered structure of $\alpha$-\ce{RuCl3} viewed along [010], f) [\ce{RuCl3}]\textsuperscript{x-} single layer on 270 nm \ce{SiO2}/Si substrate, imaged by optical microscopy, g) AFM image of highlighted [\ce{RuCl3}]\textsuperscript{x-} monolayer from optical image with 1.69(9)~nm height.}
		\label{fig:1}
\end{figure}

For the synthesis of t-\ce{RuCl3}, large single crystals of $\alpha$-\ce{RuCl3} were prepared by chemical vapor transport from 1023~K to 973~K. X-ray diffraction on a single crystal confirmed the recently reported monoclinic space group \textit{C}2/\textit{m}\cite{RuCl3-monoclinic}, in which the $\alpha$-\ce{RuCl3} layers consist of edge sharing RuCl$_6$ octahedra and are separated by a van der Waals gap. Every third octahedron is vacant, resulting in a honeycomb lattice of Ru\textsuperscript{3+} centers. The crystal structure is shown in figure \ref{fig:1} d) and e). The crystals were treated with \ce{LiBH4} in THF under inert atmosphere to yield \ce{Li_{0.2}RuCl3} according to a modified literature procedure \cite{RuCl3-intercal-Li+polymer}. Energy dispersive X-ray (EDX) and atomic emission spectroscopy confirmed the composition of \ce{Li_{0.2}RuCl3}. In contact with \ce{H2O}, \ce{Li_{0.2}RuCl3} spontaneously starts to exfoliate into $[$\ce{RuCl3}$]$\textsuperscript{x-} single layers. After shaking for 24~h, the black suspension was centrifuged to separate unreacted crystallites and larger agglomerates from the liquid. The dispersion was further washed three times by centrifugation and replacement of the supernatant liquid with water. The nanosheets form a stable suspension in water as indicated by the zeta potential of -51.4~mV. $[$\ce{RuCl3}$]$\textsuperscript{x-} nanosheets were both characterized as monolayers on 270~nm \ce{SiO2}/Si substrates by optical and atomic force microscopy (AFM) and on lacey carbon grids by transmission electron microscopy (TEM). Pellets of $[$\ce{RuCl3}$]$\textsuperscript{x-} nanosheets were prepared by evaporation of the solvent and were characterized by powder X-ray diffraction (PXRD). Monolayers on substrates and pellets of $[$\ce{RuCl3}$]$\textsuperscript{x-} were oxidized in an atomsphere of \ce{Br2} over night to yield t-\ce{RuCl3}, which was characterized by AFM and PXRD. All compounds were examined by EDX, confirming the Ru~:~Cl ratio of 1~:~3.

Figures \ref{fig:1} a) to c) show the images of $\alpha$-\ce{RuCl3} crystals, the intermediate product \ce{Li_{0.2}RuCl3} and an aqueous $[$\ce{RuCl3}$]$\textsuperscript{x-} dispersion as well as the crystal structure of $\alpha$-\ce{RuCl3}. As seen in figure~\ref{fig:1} f), monolayers of $[$\ce{RuCl3}$]$\textsuperscript{x-} were located by the enhanced interference contrast on the \ce{SiO2}/Si substrate, a method known from investigations on graphene and \ce{MoS2} \cite{graphene-how-to,TMD-sheet-visibility}. Different batches of $[$\ce{RuCl3}$]$\textsuperscript{x-} dispersions were analyzed to ensure reproducability. The single layers covered areas of up to 450~$\mu$m\textsuperscript{2} and thus their areas are of the same order of magnitude as the largest chemically exfoliated \ce{MoS2} monolayers known to date\cite{MoS2-high-area}.

\begin{figure}[thbp]
	\centering
		\includegraphics[width=0.36\textwidth]{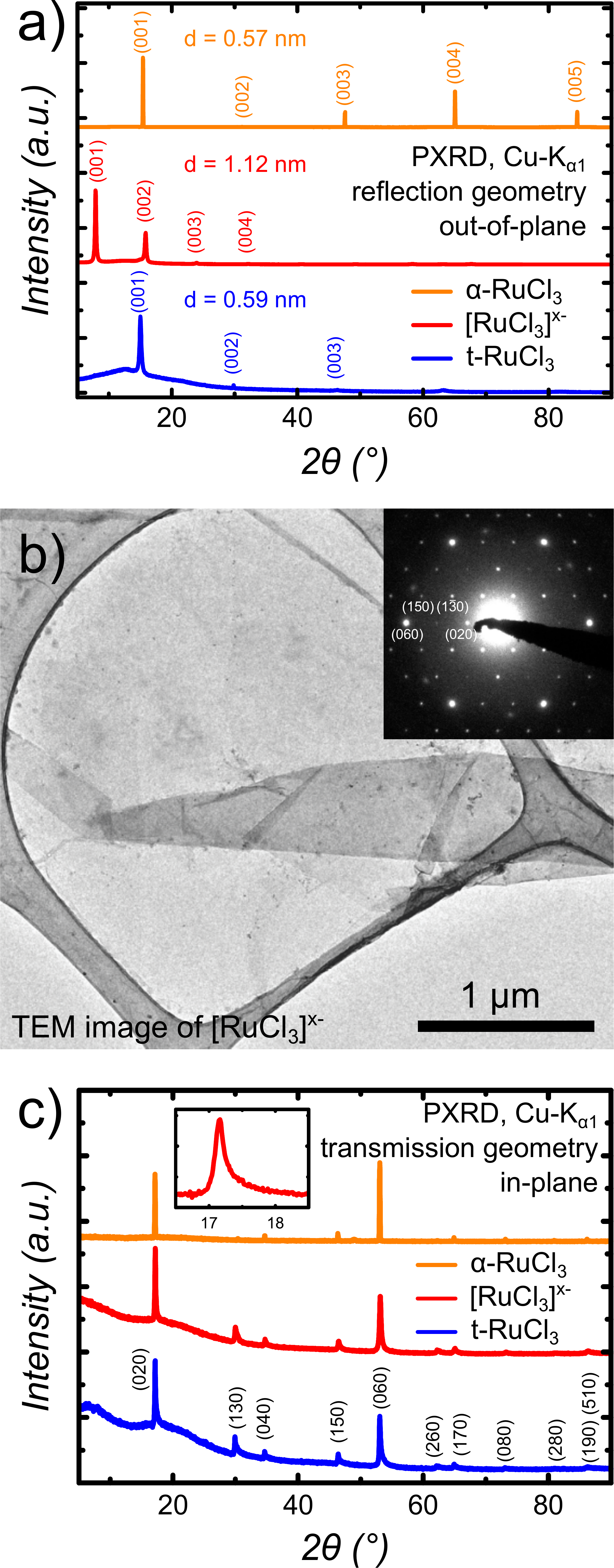}
		\caption{a) In-plane PXRD of $\alpha$-\ce{RuCl3} single crystals, [\ce{RuCl3}]\textsuperscript{x-} and t-\ce{RuCl3} nanosheet pellets measured in reflection geometry, b) TEM bright field image of partially folded [\ce{RuCl3}]\textsuperscript{x-} on lacey carbon grid, with SAED pattern of [\ce{RuCl3}]\textsuperscript{x-} as an inset, viewed along the [001] zone axis, c) in-plane PXRD of $\alpha$-\ce{RuCl3} single crystals, [\ce{RuCl3}]\textsuperscript{x-} and t-\ce{RuCl3} nanosheet pellets measured in transmission geometry, inset features the (020) peak of [\ce{RuCl3}]\textsuperscript{x-} with a Warren-type shape due to turbostratic disorder.}
		\label{fig:2}
\end{figure}

The height of the $[$\ce{RuCl3}$]$\textsuperscript{x-} single layers was analyzed by AFM for the monolayer on the \ce{SiO2}/Si substrate and by PXRD for the restacked pellet. Figure~\ref{fig:1} e) shows one exemplary AFM image with a height of 1.69(9)~nm for a $[$\ce{RuCl3}$]$\textsuperscript{x-} monolayer. It is a common phenomenon that the monolayer height on the substrate differs from the nanosheet terrace height on top of another nanosheet\cite{graphene-on-sio2}. The height of a $[$\ce{RuCl3}$]$\textsuperscript{x-} single layer folded onto itself was determined to be 1.06(6)~nm. This value is similar to that of chemically exfoliated \ce{MoS2}, where the single layer has a height of 1.0~-~1.2~nm \cite{MoS2-chem-exfol}. Out-of-plane PXRD measurements of the $[$\ce{RuCl3}$]$\textsuperscript{x-} pellet (fig. \ref{fig:2} a) in reflection geometry yielded a height of d$_{(001)}$~=~1.12~nm, confirming the results from AFM. The small height difference might originate from variations in relative humidity, which recently were shown to have a large influence on the layer height of phosphatoantimonic acid \ce{H3Sb3P2O14} nanosheets \cite{sheet-bragg-stack}.

Though the height of a single layer from PXRD and AFM are in agreement, there is still a discrepancy compared to the height of a single layer of the $\alpha$-\ce{RuCl3} crystal structure (0.57~nm). This is most likely due to the presence of residual charge on a single $[$\ce{RuCl3}$]$\textsuperscript{x-} layer, which is indicated by the high zeta potential of -~51.4~mV of the $[$\ce{RuCl3}$]$\textsuperscript{x-} dispersion. The surface charge would attract hydrated counterions, thus increasing the layer height depending on the height by the amount of water surrounding the ions. After the reductive intercalation, the ratio of Ru~:~Cl~$=$~1~:~3 remains unchanged according to EDX. Therefore, we assume that the charge injected by \ce{LiBH4} resides in the d-bands of Ru, leading to a 4d\textsuperscript{5+x} state of \ce{Ru^{(3-x)+}} in $[$\ce{RuCl3}$]$\textsuperscript{x-}.

The $[$\ce{RuCl3}$]$\textsuperscript{x-} monolayers and pellets were oxidized in a \ce{Br2} atmosphere to remove the residual layer charge, leading to t-\ce{RuCl3}. The single layer height measured by AFM shrank from 1.69(9)~nm to 1.08(17)~nm for a nanosheet on the substrate and from 1.06(6)~nm to 0.72(11)~nm in the folded monolayer. According to PXRD measurements, the restacked layers in the pellet feature a similar decrease of the stacking distance from 1.12~nm before to 0.59~nm after oxidation. The latter value is close to the layer height of 0.57~nm in bulk $\alpha$-\ce{RuCl3}. Again, the AFM and PXRD data are in good agreement with a slightly higher value from AFM, possibly due to surface roughness. Hence the present interlayer species in $[$\ce{RuCl3}$]$\textsuperscript{x-} were expelled by the oxidative treatment with \ce{Br2} to yield the oxidized nanosheets and HBr.

Since the exfoliation process induces physcial stress and bromine's high oxidative power could potentially damage the sample, the integrity of the in-plane crystal structure was investigated by TEM on the $[$\ce{RuCl3}$]$\textsuperscript{x-} single layer as well as PXRD on a multilayer stack of $[$\ce{RuCl3}$]$\textsuperscript{x-} and t-\ce{RuCl3}. Figure~\ref{fig:2} b) displays a TEM bright field image of a single $[$\ce{RuCl3}$]$\textsuperscript{x-} layer partially folded onto itself on a lacey carbon grid. The inset features the in-plane (\textit{hk}0) reflections from selected area electron diffraction (SAED). The reflections were assigned by simulationg the diffraction pattern based on the space group \textit{C}2/\textit{m} of bulk $\alpha$-\ce{RuCl3}. Figure S1 shows the simulated diffraction pattern. The \textit{d}-values of the first five most intense reflections, as listed in Table~S4, coincide with the \textit{d}-values from the simulation based on the single crystal data within the margin of error in TEM. Therefore, we conclude that the in-plane structure is maintained in the exfoliation process.

Additionally, PXRD was performed on the pellet of restacked $[$\ce{RuCl3}$]$\textsuperscript{x-} layers and t-\ce{RuCl3} in transmission geometry, to check the effect of the \ce{Br2} treatment on the in-plane structure. The resulting diffraction patterns are shown in figure~\ref{fig:2} c) and consist of the (\textit{hk}0) reflections with an intensity tail towards higher angles. The peak positions and \textit{d}-values coincide with those of the (\textit{hk}0) reflections of single crystal $\alpha$-\ce{RuCl3} (Tab. S4), indicating the retention of the in-plane structure in the $[$\ce{RuCl3}$]$\textsuperscript{x-} pellet, as well as in oxidized t-\ce{RuCl3}.

The diffraction patterns also offer information about the ordering of the layers in the multistack. A noticable feature is the Warren-type peak shape tailing off towards higher angles. The anisotropic form originates from the diffraction of the X-ray beam by a lattice with two dimensional translation symmetry without any ordering in the third dimension \cite{2D-xrd,turbostrat-disorder,ws2-turbostrat}. Therefore, the peak shape is the first indicator for the absence of order in the third dimension, signifying turbostratic disorder. A similar conclusion can be drawn from the out-of-plane PXRD, where the pellets of $[$\ce{RuCl3}$]$\textsuperscript{x-} and t-\ce{RuCl3} feature an exponential intensity decay for the series of (00\textit{l}) reflections with higher order. This is known from tetrabutylammonium (TBA) intercalated, swollen lamellar phases such as \ce{TBA_{0.35}Ti_{0.91}O2}, \ce{TBA_{0.13}MnO2} and \ce{TBA_{0.2}RuO_{2.1}}\cite{tba-titanate,tba-manganate,tba-ruthenate}. The (00\textit{l}) reflections of the ordered bulk $\alpha$-\ce{RuCl3} features a different intensity distribution, which is displayed in figure S2. Therefore, the presence of turbostratic disorder is suggested by the in- and out-of-plane PXRD patterns.

\begin{figure}[thbp]
	\centering
		\includegraphics[width=0.8\textwidth]{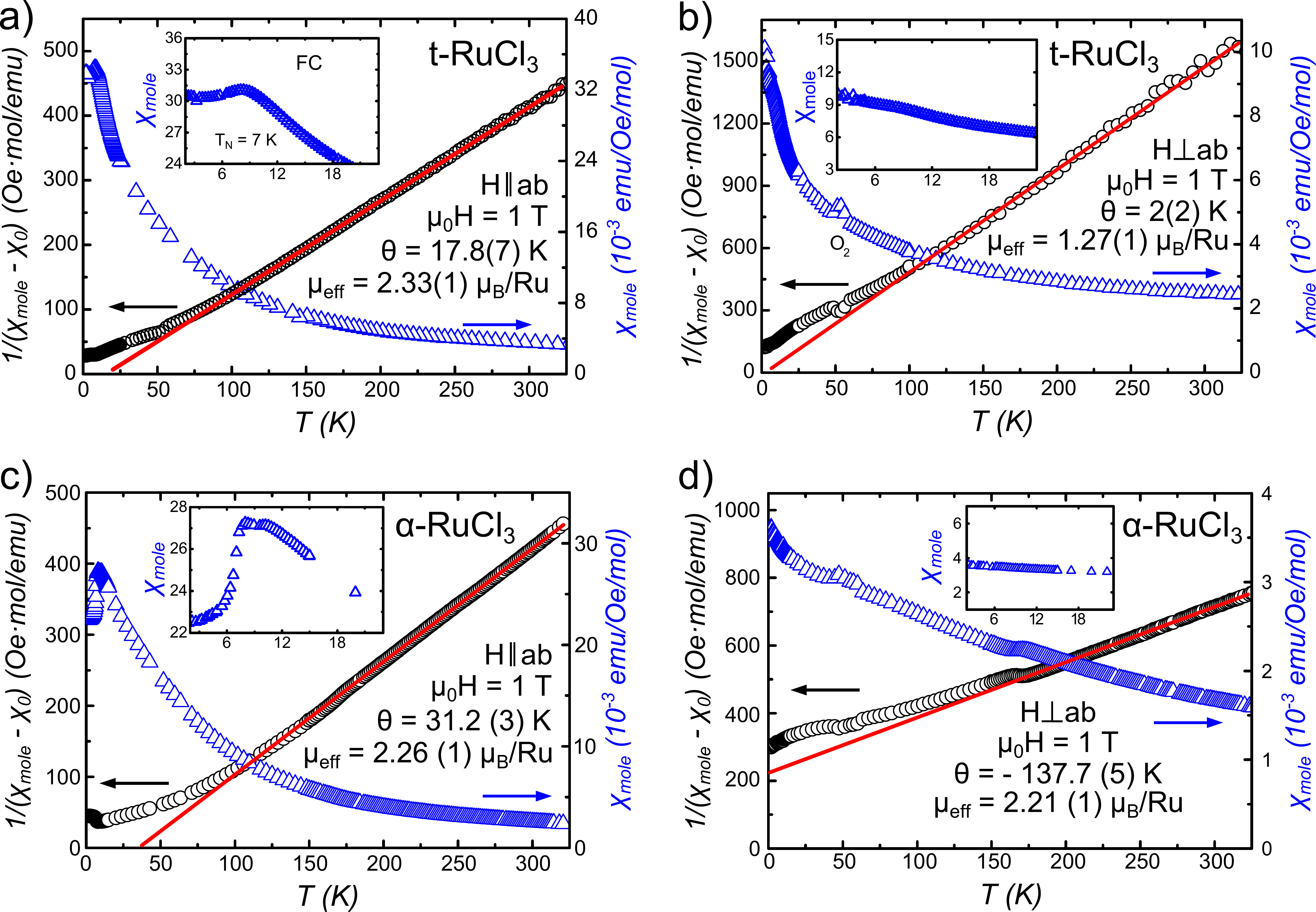}
		\caption{a) In-plane susceptibility (blue) and inverse susceptibility (black) of t-\ce{RuCl3} nanosheet pellet vs temperature at a field of $\mu_0$H~=~1~T, Curie-Weiss fit as red line, inset shows susceptiblity from T~=~2~K to 23~K with broad peak at T$_N$~=~7~K, b) out-of-plane data for the same samples measured in a field of $\mu_0$H~=~1~T, inset displays relevant region with no visible peaks in the same scale as a),
c) in-plane susceptibility (blue) and inverse susceptibility (black) of $\alpha$-\ce{RuCl3} crystal vs temperature at a field of $\mu_0$H~=~1~T, Curie-Weiss fit as red line, inset shows susceptibility from T~=~2~K to 23~K with magnetic transitions at T$_{N1}$~=~7~K and T$_{N2}$~=~13~K, b) out-of-plane data for the same $\alpha$-\ce{RuCl3} crystal measured in a field of $\mu_0$H~=~1~T, inset displays relevant region between 2 and 23~K.}
		\label{fig:3}
\end{figure}

\begin{figure}[thbp]
	\centering
		\includegraphics[width=0.9\textwidth]{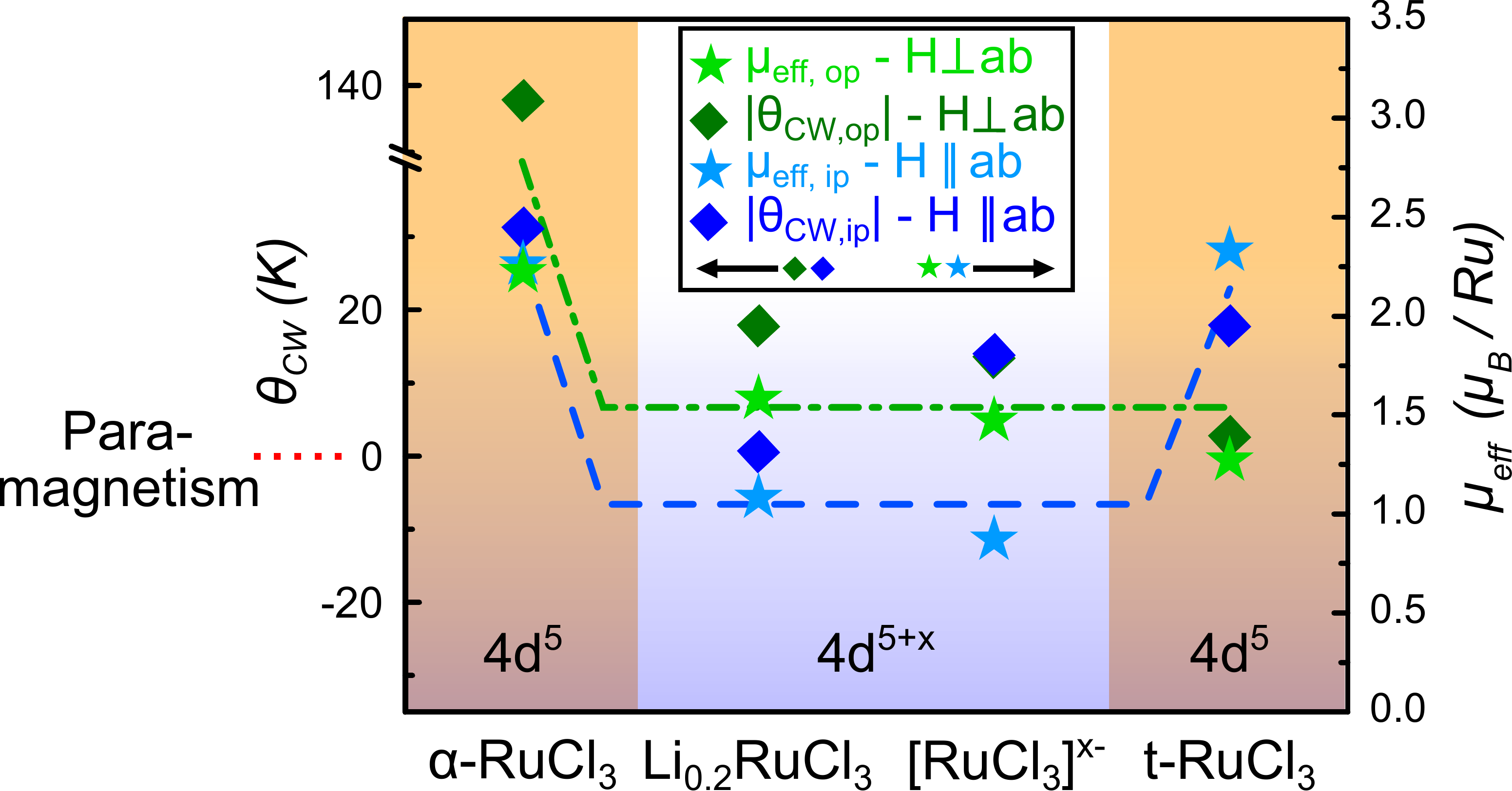}
		\caption{Evolution of the Weiss temperature $\theta_{CW}$ and effective magnetic moment $\mu_{eff}$ in-plane (blue) and out-of-plane (green) for $\alpha$-\ce{RuCl3}, \ce{Li_{0.2}RuCl3}, [\ce{RuCl3}]\textsuperscript{x-}, and t-\ce{RuCl3}. The dashed lines indicate the general trend between cooperative and non cooperative magnetism, which depends on the electronic state. $\theta_{CW}$ values were plotted as absolute values to ease comparability.}
		\label{fig:4}
\end{figure}

The introduction of turbostratic disorder and the changes in the oxidation state are expected to affect the magnetic properties of the different compounds. This is reflected in the in-plane (ip) and out-of-plane (op) investigations of the magnetic properties of $\alpha$-\ce{RuCl3} and \ce{Li_{0.2}RuCl3} as well as of the pellets of $[$\ce{RuCl3}$]$\textsuperscript{x-} and t-\ce{RuCl3}. The presence of magnetic transitions, the Weiss temperature and the magnetic moment were used as a measure for the cooperative character of the magnetic properties. The magnetic susceptibilities of t-\ce{RuCl3} and $\alpha$-\ce{RuCl3} are presented in figure \ref{fig:3}, while the data for \ce{Li_{0.2}RuCl3} and $[$\ce{RuCl3}$]$\textsuperscript{x-} is featured in the supporting information. Figure \ref{fig:4} summarizes the results of Curie-Weiss fits for all compounds.

In bulk $\alpha$-\ce{RuCl3}, we observe two magnetic transitions at \ce{T_{N1}}~=~7~K and \ce{T_{N2}}~=~13~K for the in-plane measurements, which were determined from the dMT/dT plot displayed in the supporting information and are consistent with previous experiments\cite{RuCl3-zigzag-exp,RuCl3-XY-paper,RuCl3-mag+NMR,RuCl3-QSL-proximate}. The Weiss temperatures $\theta_{CW,ip}$ of 31.2 (3) K and $\theta_{CW,op}$ of -~137.7(5)~K suggest an in-plane FM exchange and out-of-plane AF interactions. These results are comparable to previous studies, where the values range from $\theta_{CW,ip}$~=~37~K to 68~K and $\theta_{CW,op}$~=~-~145~K to -~150~K\cite{RuCl3-zigzag-exp,RuCl3-XY-paper}. The effective magnetic moment $\mu_{eff,ip}$~=~2.26(1)~$\mu_B$~/~Ru and $\mu_{eff,op}$~=~2.22(1)~$\mu_B$~/~Ru are also in the range of previously reported values ($\mu_{eff,ip}$~=~2.0~-~2.14~$\mu_B$~/~Ru and $\mu_{eff,op}$~=~2.3~-~2.7~$\mu_B$~/~Ru\cite{RuCl3-zigzag-exp,RuCl3-XY-paper}) and are much higher than the spin-only value of 1.75~$\mu_B$~/~Ru, thereby indicating the presence of SOC\cite{RuCl3-zigzag-exp,RuCl3-mag+NMR}.

Upon the reductive intercalation of lithium ions into the interlayer space, the 4d\textsuperscript{5} electron configuration of Ru\textsuperscript{3+} changes to a 4d\textsuperscript{6} state with S~=~0 for roughly 20~\% of the Ru centers in \ce{Li_{0.2}RuCl3}. This is abbreviated as "4d\textsuperscript{5.2}" in Figure \ref{fig:4}. The disturbance of the spin~$\frac{1}{2}$ order leads to paramagnetic behavior, which is associated with a decaying magnetic susceptibility with increasing temperature without any magnetic transition in the in- and out-of-plane direction. The decrease of the in-plane ($\theta_{CW,ip}$~=~0.6(2)~K, $\mu_{eff,ip}$~=~1.08(2)~$\mu_B$~/~Ru) as well as out-of-plane Weiss temperatures and magnetic moments ($\theta_{CW,op}$~=~17.8(3)~K, $\mu_{eff,op}$~=~1.58(1)~$\mu_B$~/~Ru) also reflects this trend. This suggest that the cooperative magnetism of $\alpha$-\ce{RuCl3} has been disturbed by electron injection into the \ce{RuCl3}-layers. Recently, a similar change in magnetism has been reported in \ce{Na2IrO3}, where holes were injected into the $[$\ce{Na_{1/3}Ir_{2/3}O2}$]$\textsuperscript{$\frac{2}{3}-$} layer by oxidation with \ce{Br2}. There, the low spin electron configuration changes from 5d\textsuperscript{5} to 5d\textsuperscript{4} with S~=~0 due to SOC, inducing paramagnetic behavior\cite{nairo3-paramagnet}.

The magnetic data of the restacked pellet of $[$\ce{RuCl3}$]$\textsuperscript{x-} measured within the plane exhibits a similarly decaying magnetization curve without any features, similar to the paramagnetic behavior of \ce{Li_{0.2}RuCl3}. Although the Weiss temperatures indicate slightly antiferromagnetic behavior ($\theta_{CW,ip}$~=~-~13.5(7)~K and $\theta_{CW,op}$~=~-~13.5(3)~K), no magnetic transitions were observed in the out-of-plane direction and only a slight shoulder is visible in the in-plane direction. The effective magnetic moment is very similar to that of \ce{Li_{0.2}RuCl3} with values of $\mu_{eff,ip}$~=~0.87(1)~$\mu_B$~/~Ru and $\mu_{eff,op}$~=~1.47(1)~$\mu_B$~/~Ru.

Upon oxidation of the $[$\ce{RuCl3}$]$\textsuperscript{x-} pellet by \ce{Br2} to t-\ce{RuCl3} and the associated restoration of the 4d\textsuperscript{5} state, the ordered magnetism within the plane returns. This is reflected by a transition in the in-plane data of the magnetic susceptibility at T$_N$~=~7~K. Also, it is accompanied by an increase of the Weiss temperature as well as the effective magnetic moment to $\theta_{CW,ip}$~=~17.8(7)~K and $\mu_{eff,ip}$~=~2.33(1)~$\mu_B$~/~Ru respectively, values similar to those found in the in-plane data of $\alpha$-\ce{RuCl3} ($\theta_{CW,ip}$~=~31.2~(3)~K; $\mu_{eff,ip}$~=~2.26(1)~$\mu_B$~/~Ru). In contrast, the out-of-plane measurement is characterized by paramagnetic behavior with a decaying magnetic susceptibility towards higher temperatures. The out-of-plane Weiss temperature $\theta_{CW,op}$~=~2(2)~K also indicates paramagnetic behavior, similar to the low effective magnetic moment $\mu_{eff,op}$~=~1.27(1)~$\mu_B$~/~Ru, which is much closer to the value found in the paramagnetic \ce{Li_{0.2}RuCl3} ($\mu_{eff,op}$~=~1.58(1)~$\mu_B$~/~Ru) than the one found in $\alpha$-\ce{RuCl3} ($\mu_{eff,op}$~=~2.22(1)~$\mu_B$~/~Ru).

The return of cooperative magnetism at T$_N$~=~7~K as well as the changes in the Weiss temperature and magnetic moment suggest a restoration of the magnetic order based on the 4d\textsuperscript{5} electron configuration. However, in contrast to the bulk we observe only one magnetic transition in the in-plane direction. Another difference compared to bulk $\alpha$-\ce{RuCl3} is the paramagnetic behavior of the out-of-plane data, with the Weiss temperature and the effective magnetic moment being more similar to the paramagnetic \ce{Li_{0.2}RuCl3}.

\begin{figure}[thbp]
	\centering
		\includegraphics[width=0.8\textwidth]{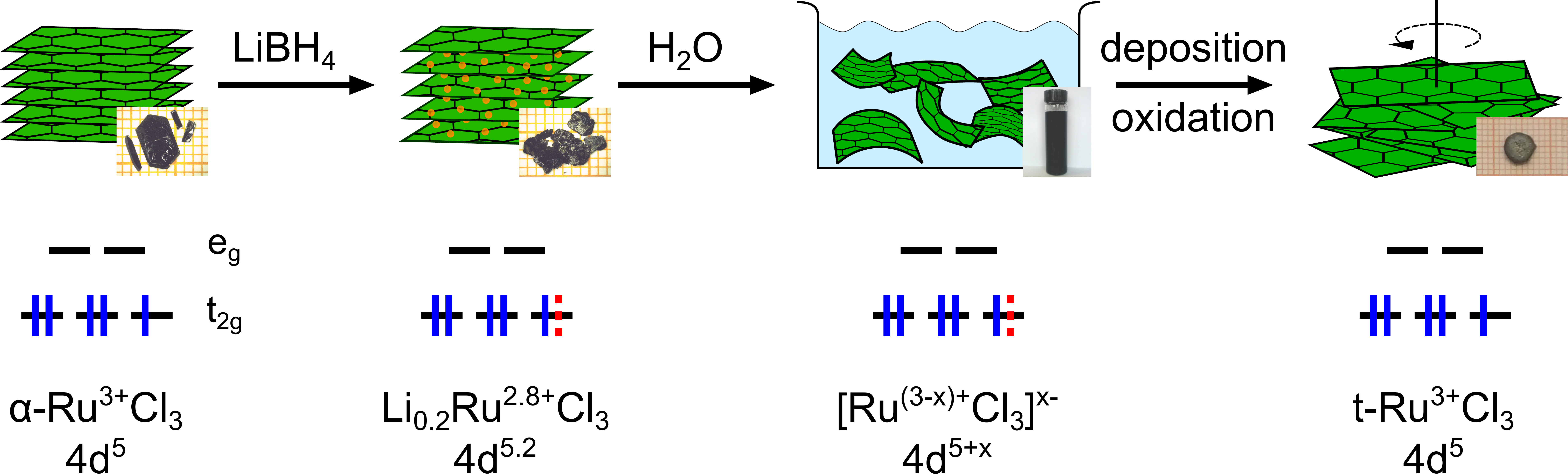}
		\caption{Schematic summary of the presented phases, the chemical steps necessary to obtain them and illustrations to explain the differences in the stacking. Simplified molecular orbitals for a monomeric $[$\ce{RuCl6}$]$\textsuperscript{3-}complex constituting a \ce{RuCl3} layer visualize the differences in the electronic state and in combination with the stacking, help to explain the evolution of the spin order.}
		\label{fig:5}
\end{figure}

To explain this behavior, the turbostratic disorder, as observed in the in- and out-of-plane PXRD data, has to be considered next to the electron configuration. Since the symmetry relation between the in-plane and the stacking direction is lifted by turbostratic disorder, no long range magnetic order can be expected outside the t-\ce{RuCl3} single layer. This could be interpreted as a structural decoupling of the nanosheets, thereby possibly weakening the magnetic interlayer interactions, which probably also affects the in-plane magnetism. Therefore, even though macroscopic pellets of \ce{RuCl3} layers were investigated, the results seem representative for the behavior of single layers of \ce{RuCl3}. Figure \ref{fig:5} gives an overview of the electronic and structural differences between the presented compounds.

Even though the exact in-plane spin structure of t-\ce{RuCl3} is yet unknown, the combination of the 4d\textsuperscript{5} electron configuration, the retention of the Ru honeycomb arrangement as well as the magnetic transition at the same temperature as the zigzag order in bulk $\alpha$-\ce{RuCl3} hint towards the restoration of the magnetic order in t-\ce{RuCl3}, with the zigzag structure being one possible candidate. 
Further investigations, especially neutron diffraction experiments, are of great interest to elucidate the spin structure of t-\ce{RuCl3} in the single layer or restacked form.

In conclusion, we presented a synthetic route towards \ce{RuCl3} nanosheets, the first exfoliation of a binary halide. Investigations of its in-plane structure show that it was retained during the exfoliation process, leading to dispersed, charged [\ce{RuCl3}]\textsuperscript{x-} monolayers in suspension. Deposition of the sheets is possible and is used to create a pellet with turbostratic disorder. Magnetic measurements show that the intermediary pellet is a paramagnet. Upon oxidizing the pellet, the long range magnetic order of the spin~$\frac{1}{2}$ honeycomb arrangement is reestablished within the plane. A very different Weiss temperature and effective magnetic moment were observed in the out-of-plane direction compared to bulk $\alpha$-\ce{RuCl3}, probably due to turbostratic disorder. Therefore, the pellet of t-\ce{RuCl3} seems to behave like a stack of magnetically decoupled single layers which can be obtained and characterized in bulk form. To our knowledge, a top-down approach of exfoliation and restacking of nanosheets to deliberately introduce turbostratic disorder enabling the manipulation of the magnetic properties of a solid has not been reported yet. We believe that the approach presented herein provides a synthetic tool to establish macroscopic quasi 2D model systems for Kitaev-Heisenberg physics in spin~$\frac{1}{2}$ honeycomb magnets and areas beyond.

\acknowledgement

The authors thank Hidenori Takagi for measurement time on the Quantum Design PPMS and Tomohiro Takayama for extremely helpful scientific discussions as well as critical reading of the manuscript. We also thank Eva Br\"ucher for preliminary magnetic measurements and Gisela Siegle for the specific heat measurements. We gratefully acknowledge the financial support by the Max Planck Society, the Nanosystems Initiative Munich (NIM) and the Center for Nanoscience (CeNS). Leslie M. Schoop gratefully acknowledges financial support by the Minerva fast track fellowship.

\begin{suppinfo}

Detailed experimental procedures, techniques used for the characterization, single crystal X-ray data, further TEM images, magnetic data on all compounds as well as specific heat data of t-\ce{RuCl3} are presented in the supporting information.

\end{suppinfo}

\pagebreak
 
\pagebreak


\pagebreak
\vspace{\baselineskip}
\begin{center}
\textbf{\huge\linespread{3} Supplemental Materials - Magnetic properties of restacked 2D spin~$\frac{1}{2}$ honeycomb RuCl$_3$ nanosheets}
\end{center}
\setcounter{equation}{0}
\setcounter{figure}{0}
\setcounter{table}{0}
\setcounter{page}{1}
\makeatletter
\renewcommand{\theequation}{S\arabic{equation}}
\renewcommand{\thefigure}{S\arabic{figure}}
\renewcommand{\bibnumfmt}[1]{[S#1]}
\renewcommand{\citenumfont}[1]{S#1}

\section{Synthesis}

\paragraph{$\alpha$-\ce{RuCl3}}
Large crystals were grown by chemical vapor transport as described in the literature\cite{RuCl3-cvt}. Commercial \ce{RuCl3} powder (99.9 \%, Roth) was sealed under vacuum in a quartz ampoule. The powder was heated with 3 K min\textsuperscript{-1} to 1023 K and held there for 36~h to 120~h, depending on the amount with a temperature gradient of approximately 50~K to 75~K from the feed to the growth zone. The reaction yielded black, crystalline platelets with edge lengths of up to 3 mm, which were analysed by means of PXRD, single crystal X-ray diffraction, SEM-EDX and elemental analysis. Smaller crystals for single crystal X-ray diffraction were grown at 923 K.

\paragraph{Li$_{0.2}$RuCl$_3$}
The synthesis is a modification of a known procedure\cite{RuCl3-intercal-Li+polymer-SI}. $\alpha$-\ce{RuCl3} was combined with tetrahydrofurane (THF) and \ce{LiBH4} under argon atmosphere. In a typical synthesis, 11.77 mL dry THF were added to 483 mg (2.33~mmol, 1 eq.) $\alpha$-\ce{RuCl3} crystals in a Schlenk flask under Ar atmosphere. 0.23 mL of \ce{LiBH4} (2 mol L\textsuperscript{-1} in THF, 0.466~mmol, 0.2~eq.) were added under counter flow of argon and the mixture was stirred over night. After washing the product with 3x 12~mL dry THF, \ce{Li_{0.2}RuCl3} was introduced into double distilled \ce{H2O} and shaken for over night to yield a black suspension. The remaining solid was separated by centrifugation. The concentration of the dispersion was 2.3~mM. The dispersion was washed by centrifugation at 18000~rpm and the remaining clear supernatant separated from the pellet. The solid pellet was redispersed in water and the washing procedure repeated three times. Pellets were prepared by dropping parts of the dispersion onto a glass substrate. Upon solvent evaporation, new drops were added until the desired amount was reached.

\paragraph{t-\ce{RuCl3}}
$[$\ce{RuCl3}$]$\textsuperscript{x-} pellets on substrates were placed in a closed vessel with some droplets of bromine over night.

\section{Single crystal X-ray diffraction}

\begin{table}[hb]
	\caption{Crystal data and structure refinement data of $\alpha$-\ce{RuCl3}.}
	\label{tab:xtal-data}
	\centering
  \begin{tabular}{l l}
		\hline
															& $\alpha$-\ce{RuCl3} \\
		\hline			
			Temperature / K 				&	298(2) K	\\
			Formula weight / g/mol\textsuperscript{-1}	& 207.43 		  \\
			Space group (no.), Z		& \textit{C} 2\textit{/m}, 4 \\
			Lattice constants / \AA & a = 5.9917(13) \\
															& b = 10.367(2) \\
															& c = 6.0543(13) \\
			V / \AA\textsuperscript{3}, $\rho_{X-ray}$ / g cm\textsuperscript{-3} &  356.21(13), 3.868 \\
			Crystal size / mm\textsuperscript{-3} & 0.10 x 0.08 x 0.04 \\
			Diffractometer					& SMART APEX I, Bruker AXS \\
			X-ray radiation, $\lambda$ / \AA & 0.71073 \\
			Absorption correction		& Multi-scan, TWINABS\cite{twinabs} \\
			2$\theta$ range / \textsuperscript{$\circ$} & 5.73 - 68.75 \\
			Index range							& -9 $\leq$ \textit{h} $\leq$ 9 \\
															& 0  $\leq$ \textit{k} $\leq$ 16 \\
															& 0  $\leq$ \textit{l} $\leq$ 9 \\
			Reflection collected		& 2803 \\
			Data, \textit{R}$_int$ & 748, 0.052 \\
			No. of parameters				& 27 \\
			Transmission: \textit{t}$_max$, \textit{t}$_min$ & 0.271, 0.167 \\
			\textit{R}$_1$$[F^2 > 2\sigma(F^2)]$ & 0.043 \\
			\textit{wR}(F\textsuperscript{2}) & 0.108 \\
			Twin volume fraction		& 0.27 \\
			$\Delta\rho$$_max$, $\Delta\rho_min$ / e \AA\textsuperscript{-3} & 2.437, -1.880 \\
		\hline
  \end{tabular}
\end{table}

Diffraction data sets were collected at 298~K on a three circle diffractometer (Bruker AXS, Karlsruhe, Germany) equipped with SMART APEX I CCD, using Mo-K$_\alpha$ radiation ($\lambda$ = 0.71073 \AA). The collection and reduction of data were carried out with the BRUKER SUITE software package\cite{bruker}. It turned out that the crystal under investigation was systematically twinned (dovetail twin for the monoclinic system), and the twin-law ( 100, 0-10, 001) had to be applied during data reduction. The intensities were corrected for absorption effects applying a multi-scan method with TWINABS\cite{twinabs}. The structure was solved by Direct Methods and refined by full matrix least-squares fitting with the SHELXTL software package\cite{shelxtl}. Some residual occupancy was found in the empty octahedron of the $\alpha$-\ce{RuCl3} structure, a consequence of disorder in the layer stacking. Experimental details of data collection and crystallographic data are given in Tables S\ref{tab:xtal-data}, S\ref{tab:xtal-pos} and S\ref{tab:xtal-adp}.

\begin{table}[htb]
	\caption{Atomic coordinates and equivalent displacement parameters \textit{U}\textsuperscript{eq} / \AA\textsuperscript{2} x 10\textsuperscript{4} for $\alpha$-\ce{RuCl3} at 298 K.}
	\label{tab:xtal-pos}
	\centering
  \begin{tabular}{l l l l l l l}
		\hline
			Atom & site & x & y & z & SOF & \textit{U}\textsuperscript{eq} \\
		\hline			
			Ru1	& 4\textit{g}	& 0	& 0.33338(4)	& 0 &	0.955(4)	& 124(2) \\
			Ru2	& 2\textit{a}	& 0	& 0	& 0 &	2-2xSOF(Ru1) & 394(44) \\
			Cl1	& 8\textit{j}	& 0.7513(2)	& 0.1736(1)	& 0.7681(2) &	1	& 174(3) \\
			Cl2	& 4\textit{i}	& 0.7277(3)	& 0	& 0.2340(3) &	1 & 181(4) \\
		\hline
  \end{tabular}
\end{table}

\begin{table}[htb]
	\caption{Anisotropic displacement parameters \textit{U}\textsuperscript{ij} / \AA\textsuperscript{2} x 10\textsuperscript{4} for $\alpha$-\ce{RuCl3} at 298 K.}
	\label{tab:xtal-adp}
	\centering
  \begin{tabular}{l l l l l l l}
		\hline
			Atom	& \textit{U}\textsuperscript{11} & \textit{U}\textsuperscript{22} & \textit{U}\textsuperscript{33} & \textit{U}\textsuperscript{12} & \textit{U}\textsuperscript{13} & \textit{U}\textsuperscript{23} \\
		\hline			
			Ru1	& 128(3)	& 107(3)	& 155(3)	& 0 &	70(2)	& 0 \\
			Ru2	& 583(91) & 367(70)	& 315(68) & 0 & 262(60) & 0 \\
			Cl1	& 180(6) & 181(5) & 191(6) & -35(3) & 101(4) & -37(3) \\
			Cl2	& 183(7) & 147(7) & 196(7) & 0 & 37(6) & 0 \\
		\hline
  \end{tabular}
\end{table}

\section{Other analytical methods}

PXRD in-plane measurements were performed on a Stoe Stadi-P (Stoe Darmstadt, Germany) utilizing Cu-K$_{\alpha1}$ radiation (Ge(111) monochromator, $\lambda$ = 1.54059 \AA) and a Mythen Dectris detector in transmission geometry. The out-of-plane measurements were performed on a Bruker D8-Advance with Cu-K$_{\alpha1}$ radiation (Ge(111) monochromator, $\lambda$ = 1.54059 \AA), in reflection geometry, using a Vantec detector. All powder diffraction patterns were recorded at room temperature.

Elemental analysis for lithium and ruthenium was performed on a Vista Pro ICP-AES spectrometer. Ruthenium to chloride ratios were measured on a scanning electron microscope (SEM; Vega TS 5130 MM, Tescan) with a SEM-EDX using a Si/Li detector (Oxford).

AFM was performed on a MFP-3D AFM by Asylum Research / Oxford Instruments in intermittent contact mode with Olympus cantilever (resonance frequency approximately 300 kHz).

TEM samples were prepared by dropping the colloidal nanosheet suspension of $[$\ce{RuCl3}$]$\textsuperscript{x-} onto a lacey carbon film/copper grid (Plano) and subsequent drying under IR-light irradiation. TEM was performed with a Phillips CM30 ST (300 kV, \ce{LaB6} cathode), with a CMOS camera (TemCam-F216, Tietz) for recording bright field images and selected area electron diffraction (SAED) patterns.

Measurements of the magnetic properties were performed on a MPMS by Quantum Design. The specific heat data was collected on a PPMS, also by Quantum Design.

\section{Further Results}

\subsection{TEM data on $[$\ce{RuCl3}$]$\textsuperscript{x-} monolayer}

\begin{figure}[thbp]
	\centering
		\includegraphics[width=0.6\textwidth]{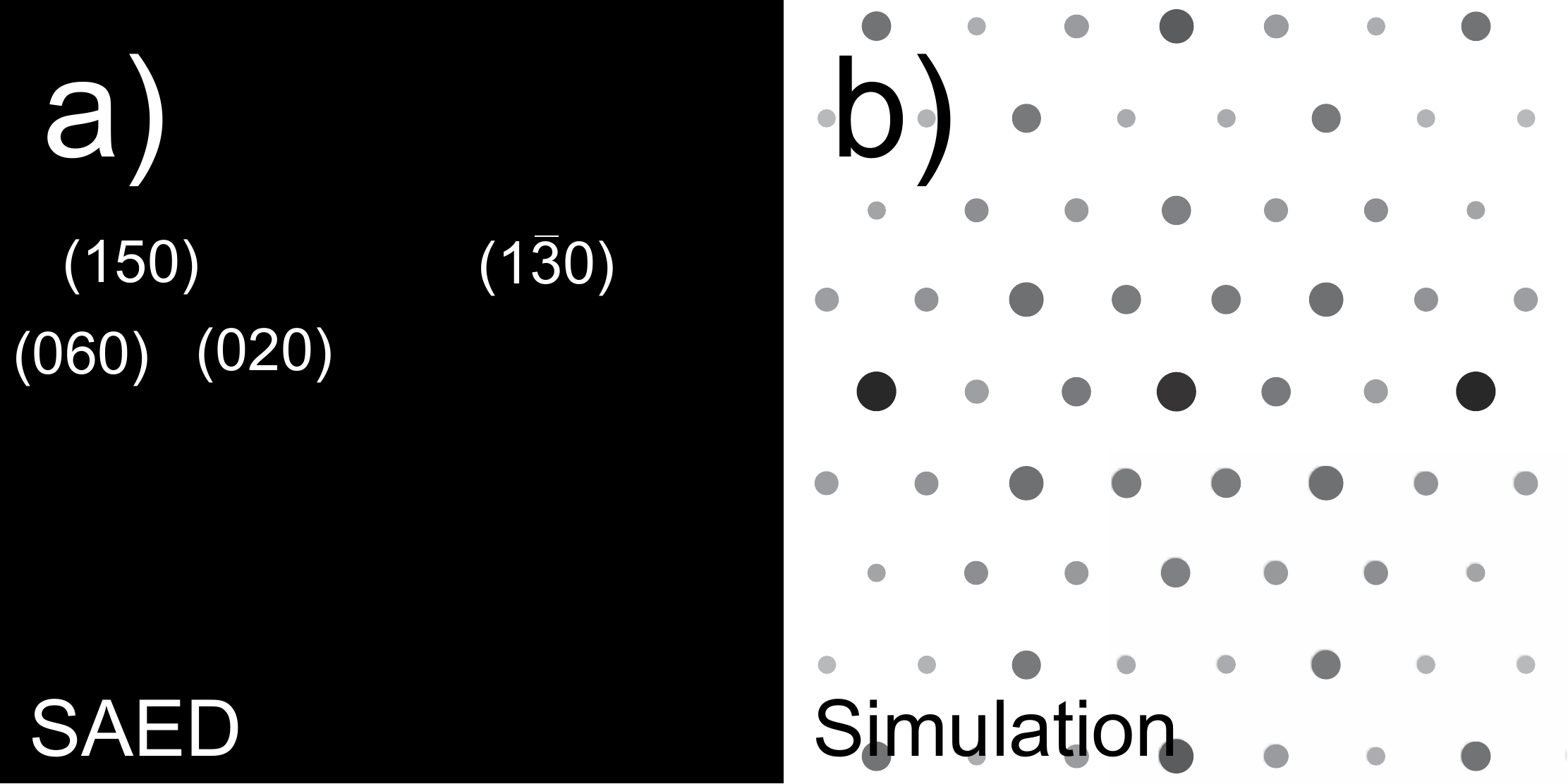}
		\caption{a) Selected area electron diffraction pattern of $[$\ce{RuCl3}$]$\textsuperscript{x-} along [001] zone axis, b) simulation of $\alpha$-\ce{RuCl3} based on the space group \textit{C}2/\textit{m} obtained from single crystal X-ray diffraction.}
		\label{fig:saed+sim}
\end{figure}

\begin{table}[htb]
	\caption{Comparison of \textit{d}-values from TEM of a $[$\ce{RuCl3}$]$\textsuperscript{x-} single layer, PXRD on the restacked pellet of $[$\ce{RuCl3}$]$\textsuperscript{x-} and single crystal X-ray diffraction on $\alpha$-\ce{RuCl3}.}
	\label{tab:d-values}
	\centering
  \begin{tabular}{c c c c}
		\hline
			(\textit{hkl}) & \textit{d}-values TEM & \textit{d}-values PXRD & \textit{d}-values \\
						& $[$\ce{RuCl3}$]$\textsuperscript{x-} & $[$\ce{RuCl3}$]$\textsuperscript{x-} & $\alpha$-\ce{RuCl3}  \\
						& [nm] 				& [nm] 			& [nm] \\
			\hline			
			(020) & 5.27 				&  5.16 		&	5.18 \\
			(130) & 3.04 				&  2.98			&	2.95 \\
			(040) & 2.64 				&  2.57			&	2.59 \\
			(150) & 1.99 				&  1.96			&	1.95 \\
			(060) & 1.76 				&  1.72			&	1.73 \\
    \hline
  \end{tabular}
\end{table}

\pagebreak

\subsection{Magnetic Data and Curie-Weiss Fits}

\begin{table}[hb]
	\caption{Weiss temperature $\theta_{CW}$ and effective magnetic moment $\mu_{eff}$ from Curie-Weiss fits for the in-plane (ip) and out-of-plane (op) direction.}
	\label{tab:CW}
	\centering
  \begin{tabular}{l c c c c }
		\hline
			Compound & $\theta_{CW,ip}$ & $\theta_{CW,op}$ & $\mu_{eff,ip}$ & $\mu_{eff,op}$ \\
			 & [K] & [K]  & [$\mu_B$ / Ru] & [$\mu_B$ / Ru] \\
		\hline			
			$\alpha$-\ce{RuCl3}	& 31.2(3) & - 137.7(5) & 2.26(1) & 2.22(1) \\
			\ce{Li_{0.2}RuCl3} & 0.6(2) & - 17.8(3) & 1.08(2) & 1.58(1) \\
			$[$\ce{RuCl3}$]$\textsuperscript{x-} & - 13.5(7) & - 13.5(3) & 0.87(1) & 1.47(1) \\
			\ce{RuCl3}-T & 17.8(7) & 2(2) & 2.33(1) & 1.27(1) \\
		\hline
  \end{tabular}
\end{table}

\begin{figure}[thbp]
	\centering
		\includegraphics[width=0.6\textwidth]{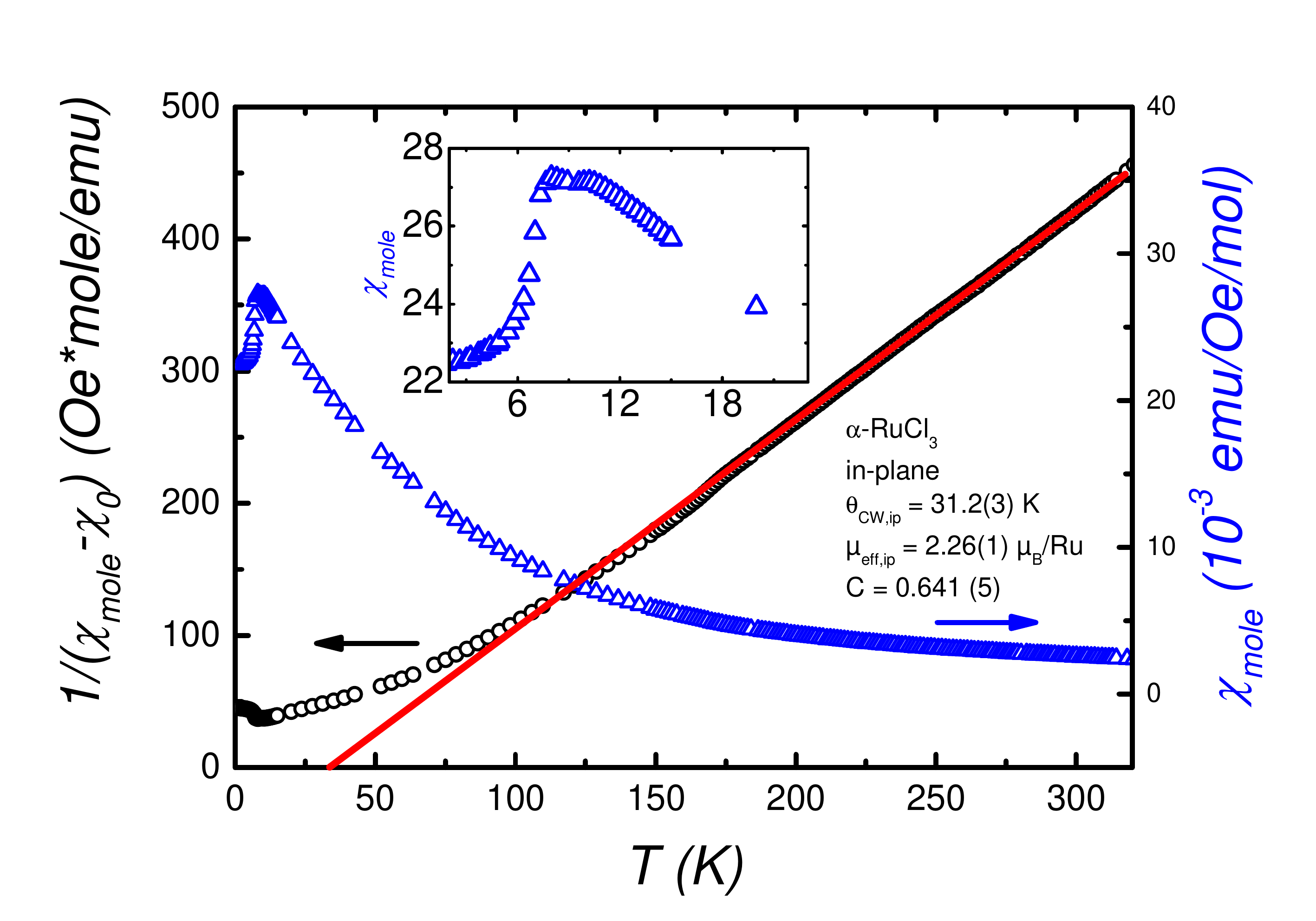}
		\caption{In-plane magnetic susceptibility and inverse magnetic susceptibility of $\alpha$-\ce{RuCl3} vs temperature at $\mu_0$H~=~1~T, inset shows susceptibility from T~=~3~K to 23~K.}
		\label{fig:alpha-RuCl3-ip}
\end{figure}

\begin{figure}[thbp]
	\centering
		\includegraphics[width=0.6\textwidth]{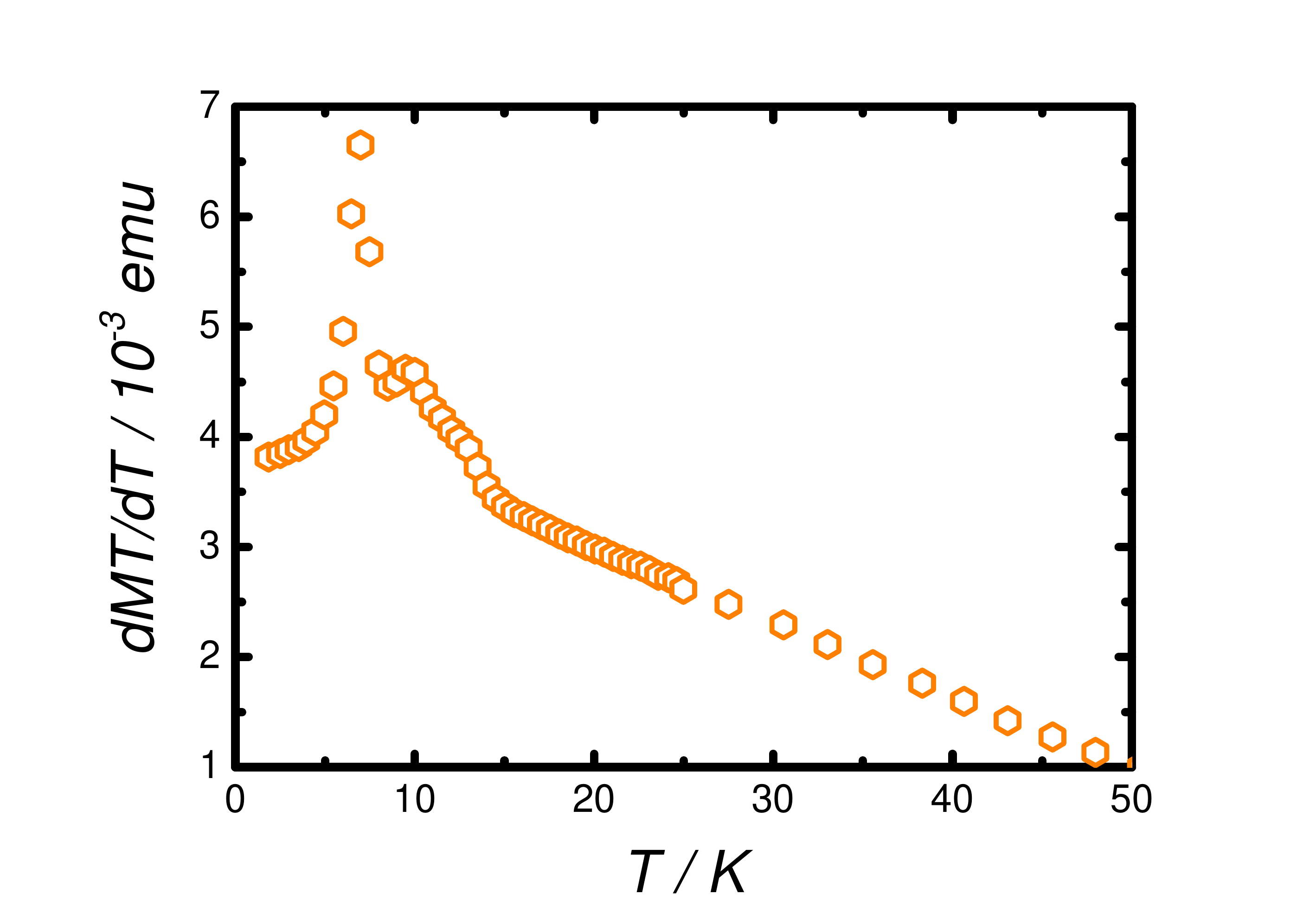}
		\caption{Plot of dMT/dT for T~=~3~K to 23~K for the in-plane measurement of $\alpha$-\ce{RuCl3}. Magnetization (M) is taken as $\chi H$.}
		\label{fig:alpha-RuCl3-dMT/dT}
\end{figure}

\begin{figure}[thbp]
	\centering
		\includegraphics[width=0.6\textwidth]{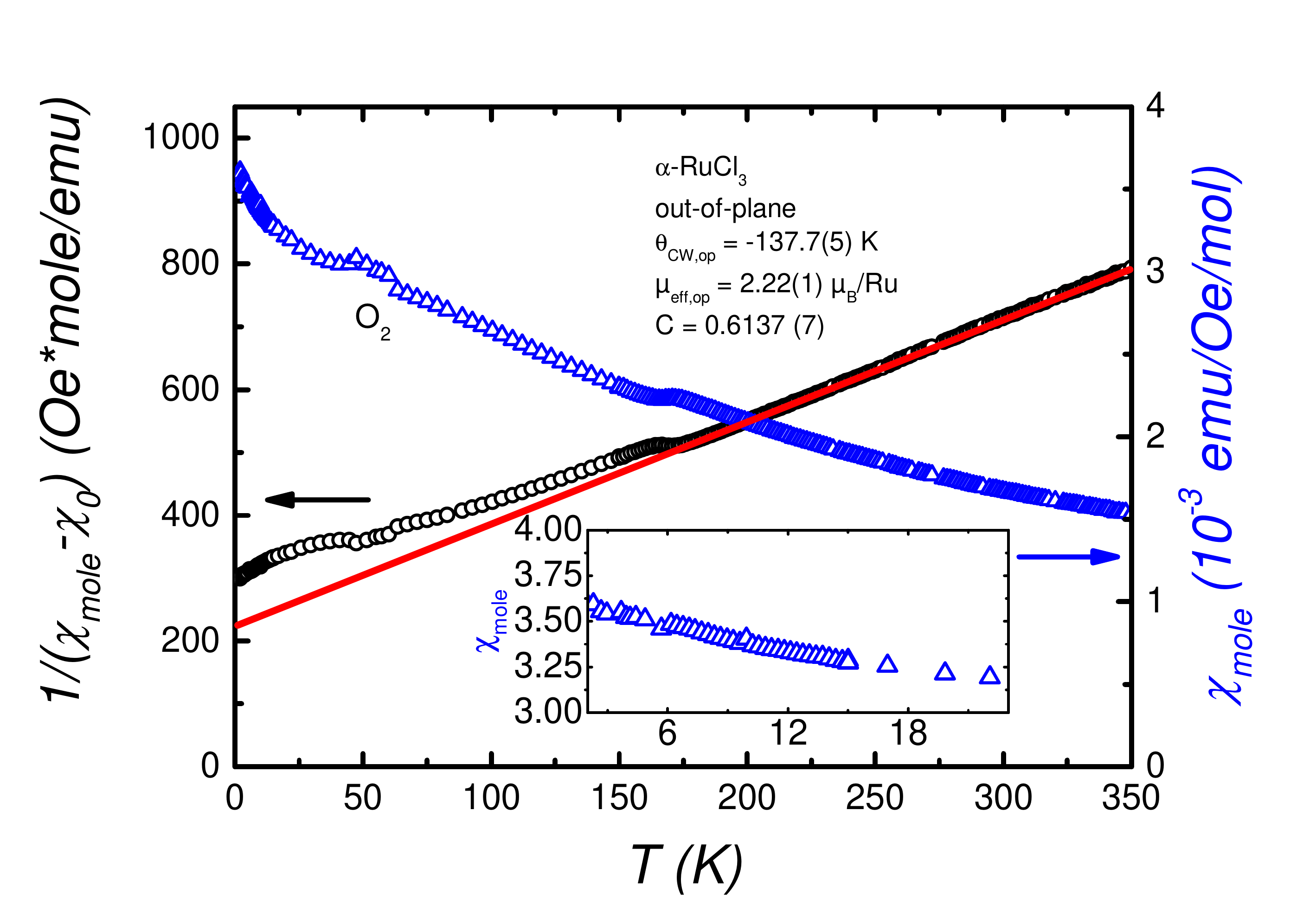}
		\caption{Out-of-plane magnetic susceptibility and inverse magnetic susceptibility of $\alpha$-\ce{RuCl3} vs temperature at $\mu_0$H~=~1~T, inset shows susceptibility from T~=~3~K to 23~K. The discontinuity at around 160 K was observed in the literature\cite{RuCl3-XY-paper-SI} and assumed to be a change in the stacking behavior resulting in a structural phase transition from \textit{C}2/\textit{m} to \textit{R}$\bar{3}$.}
		\label{fig:alpha-RuCl3-op}
\end{figure}

\begin{figure}[thbp]
	\centering
		\includegraphics[width=0.6\textwidth]{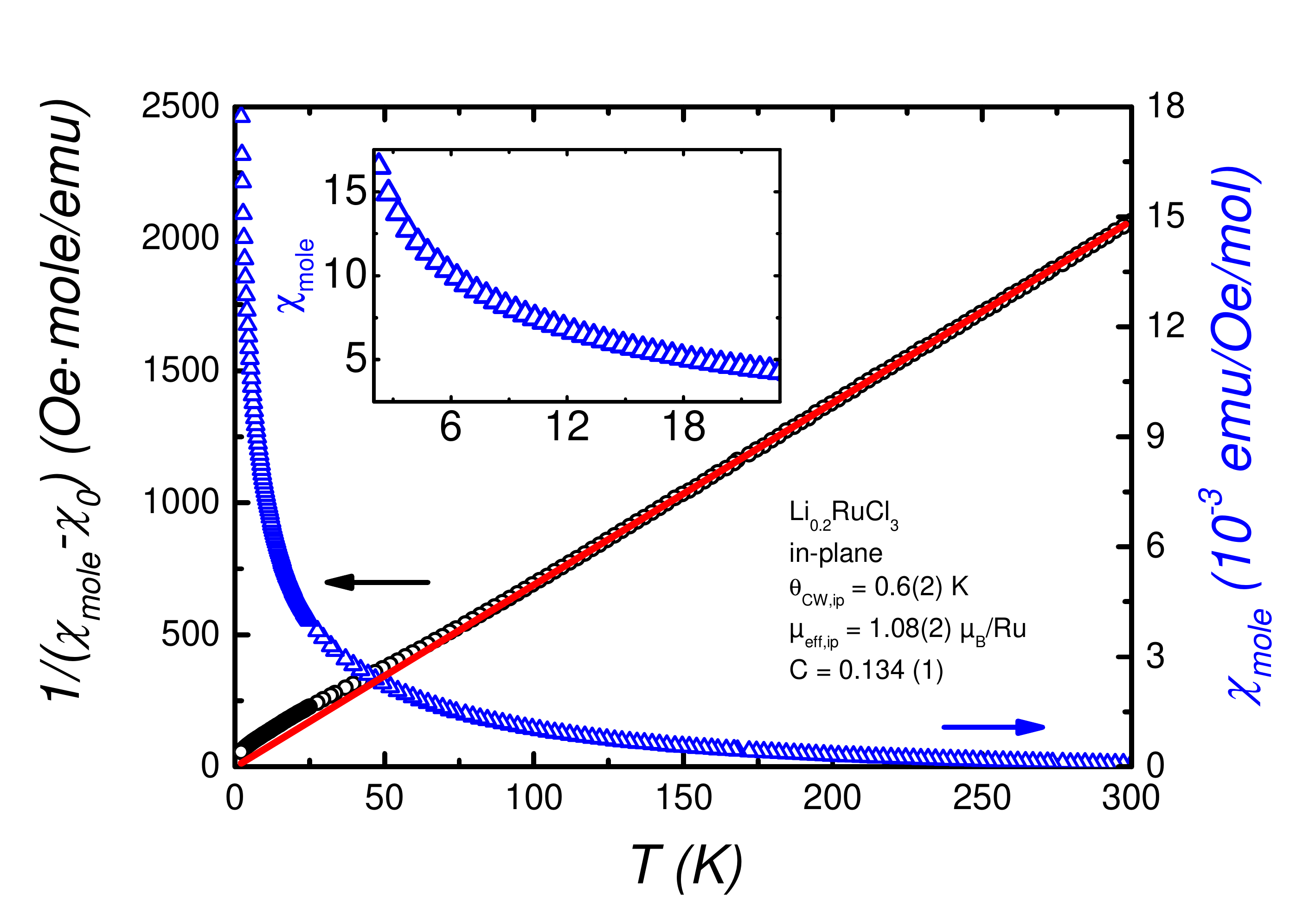}
		\caption{In-plane magnetic susceptibility and inverse magnetic susceptibility of \ce{Li_{0.2}RuCl3} vs temperature at $\mu_0$H~=~1~T, inset shows susceptibility from T~=~3~K to 23~K.}
		\label{fig:Li02RuCl3-ip}
\end{figure}

\begin{figure}[thbp]
	\centering
		\includegraphics[width=0.6\textwidth]{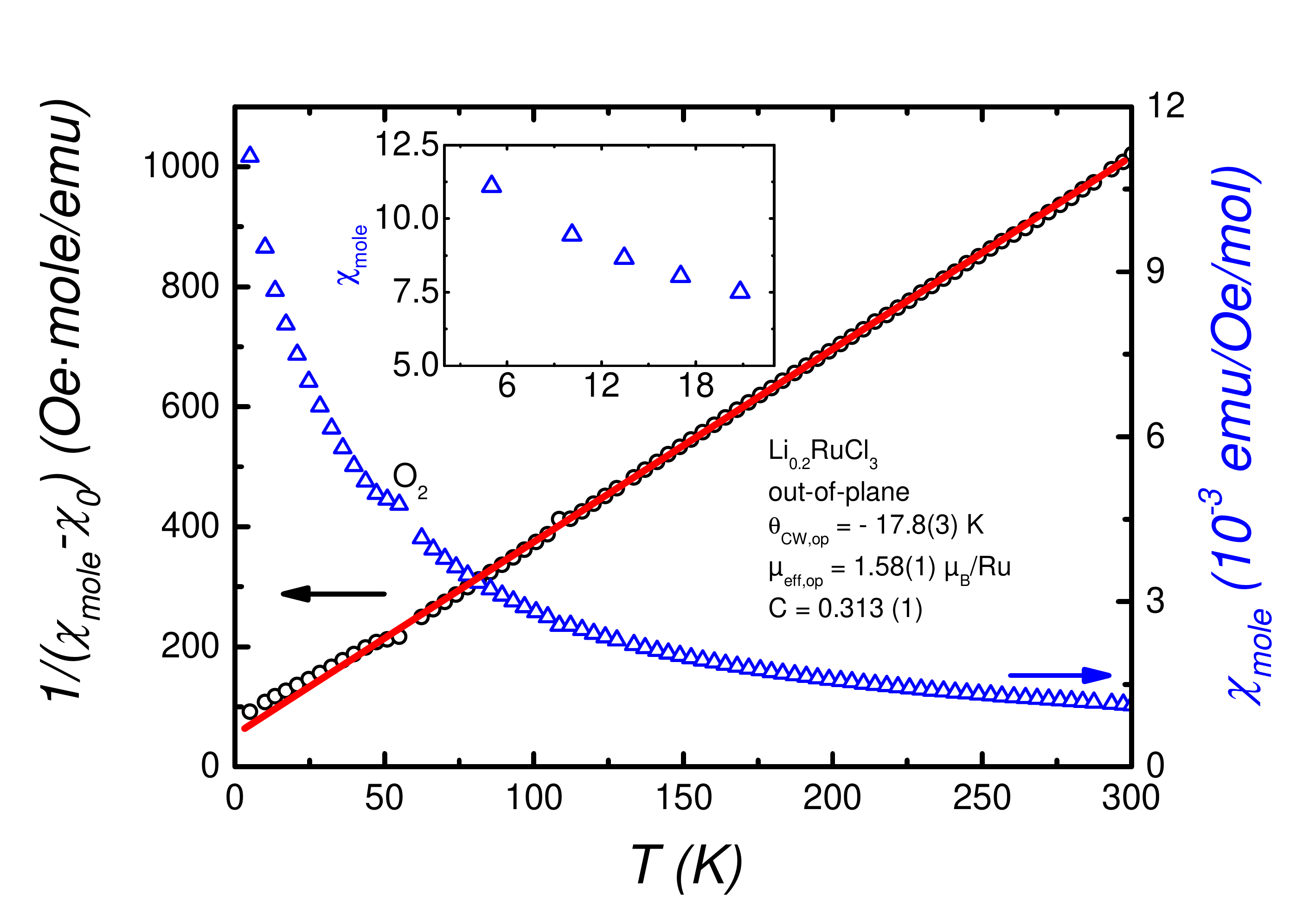}
		\caption{Out-of-plane magnetic susceptibility and inverse magnetic susceptibility of \ce{Li_{0.2}RuCl3} vs temperature at $\mu_0$H~=~1~T, inset shows susceptibility from T~=~3~K to 23~K.}
		\label{fig:Li0.2RuCl3-op}
\end{figure}

\begin{figure}[thbp]
	\centering
		\includegraphics[width=0.6\textwidth]{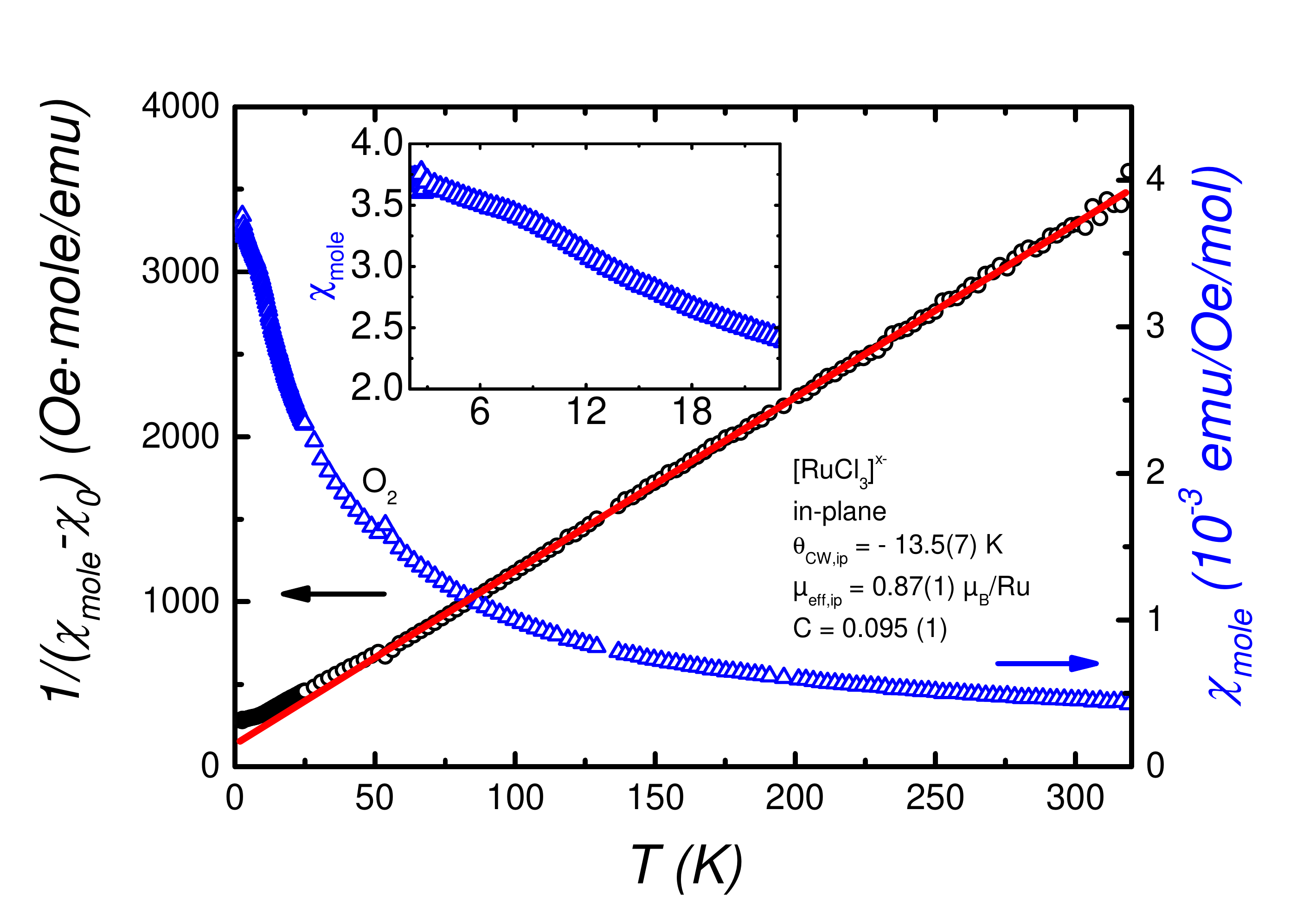}
		\caption{In-plane magnetic susceptibility and inverse magnetic susceptibility of $[$\ce{RuCl3}$]$\textsuperscript{x-} vs temperature at $\mu_0$H~=~1~T, inset shows susceptibility from T~=~3~K to 23~K.}
		\label{fig:RuCl3x-ip}
\end{figure}

\begin{figure}[thbp]
	\centering
		\includegraphics[width=0.6\textwidth]{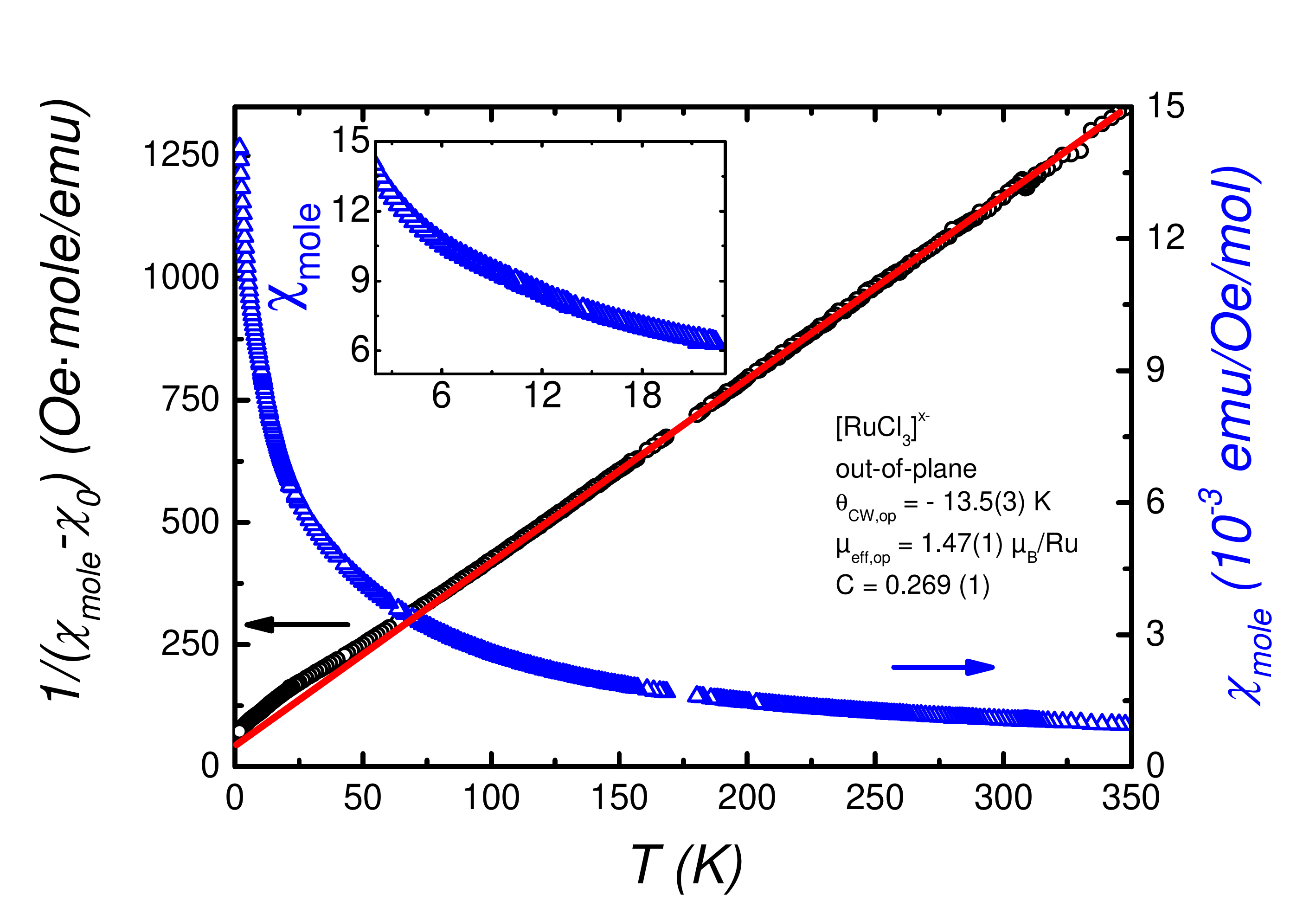}
		\caption{Out-of-plane magnetic susceptibility and inverse magnetic susceptibility of $[$\ce{RuCl3}$]$\textsuperscript{x-} vs temperature at $\mu_0$H~=~1~T, inset shows susceptibility from T~=~3~K to 23~K.}
		\label{fig:RuCl3x-op}
\end{figure}

\begin{figure}[thbp]
	\centering
		\includegraphics[width=0.6\textwidth]{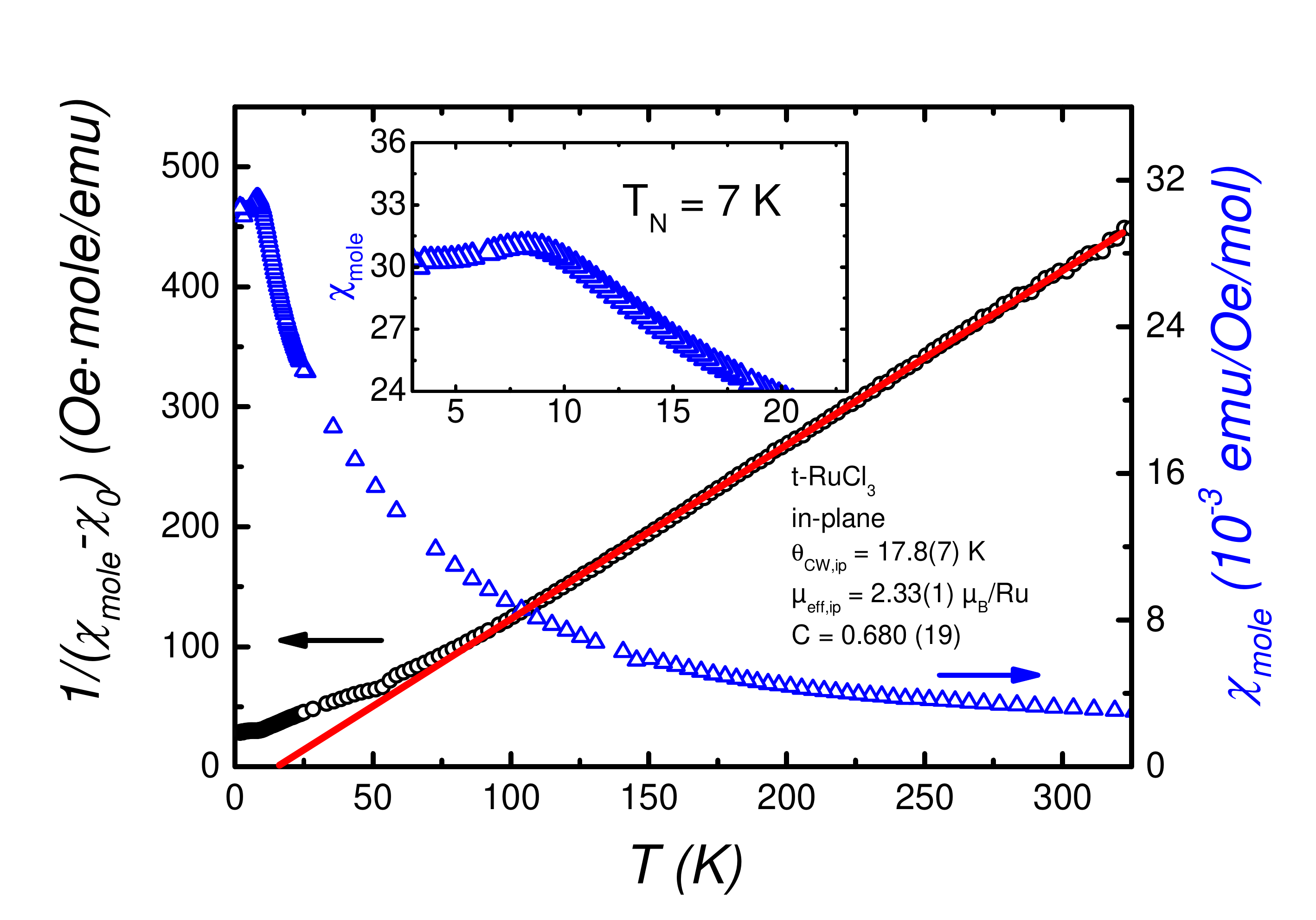}
		\caption{In-plane magnetic susceptibility and inverse magnetic susceptibility of t-\ce{RuCl3} vs temperature at $\mu_0$H~=~1~T, inset shows susceptibility from T~=~3~K to 23~K.}
		\label{fig:RuCl3-T-ip}
\end{figure}

\begin{figure}[thbp]
	\centering
		\includegraphics[width=0.6\textwidth]{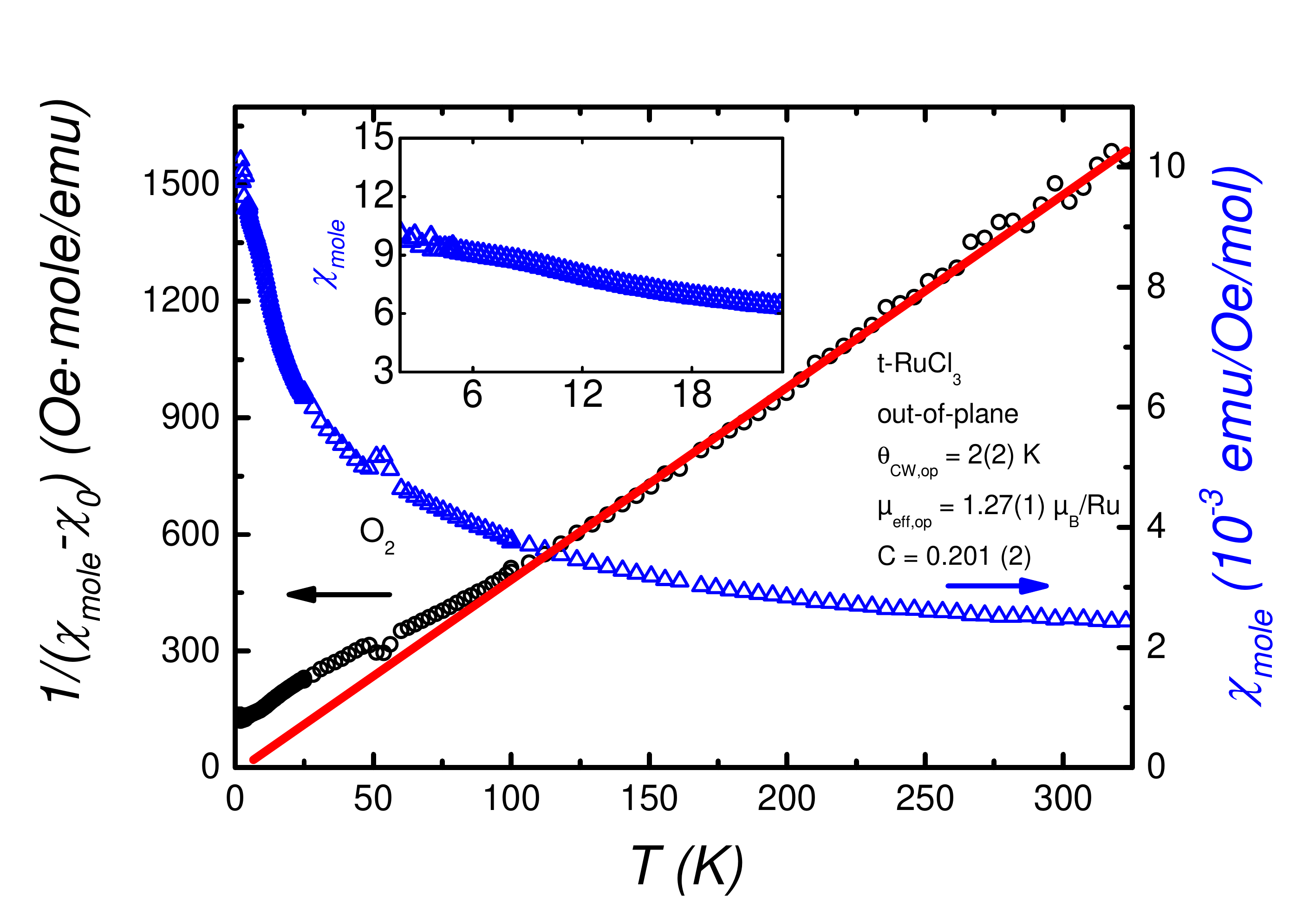}
		\caption{Out-of-plane magnetic susceptibility and inverse magnetic susceptibility of t-\ce{RuCl3} vs temperature at $\mu_0$H~=~1~T, inset shows susceptibility from T~=~3~K to 23~K.}
		\label{fig:RuCl3-T-op}
\end{figure}

\newpage

\subsection{Specific heat measurement}

\begin{figure}[thbp]
	\centering
		\includegraphics[width=0.6\textwidth]{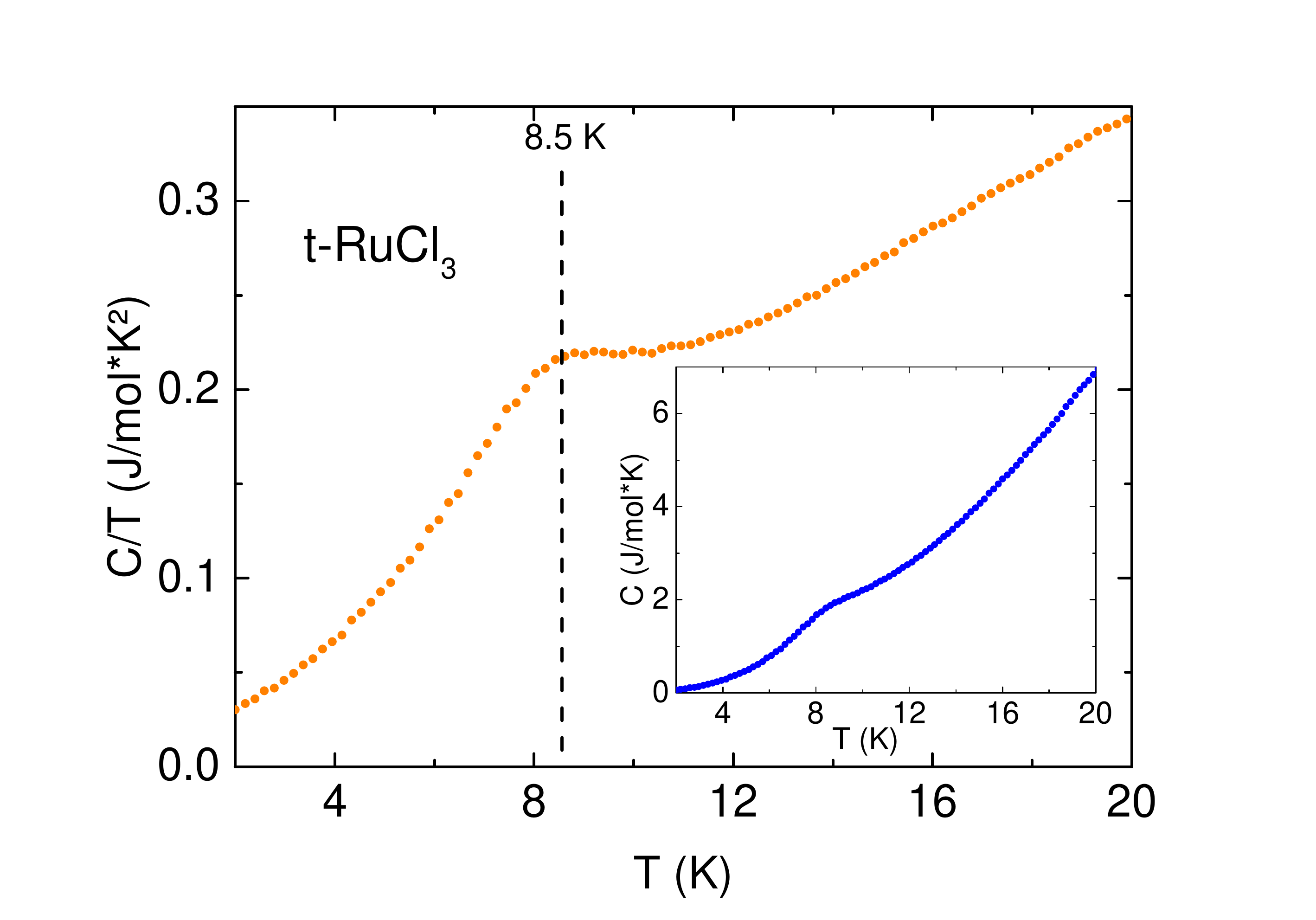}
		\caption{Temperature dependent specific heat of t-\ce{RuCl3}. A magnetic transition is visible at 8.5~K, no further transitions can be detected below 20~K.}
		\label{fig:RuCl3-HC}
\end{figure}

\end{document}